\documentclass[useAMS,usenatbib]{mnras}
\usepackage{times}
\usepackage{color}
\usepackage[fleqn]{amsmath}
\usepackage{aas_macros,amsfonts,amssymb}
\usepackage{url,hyperref}
\usepackage{mathptmx}
\usepackage{graphicx}
\usepackage{booktabs}
\usepackage{flushend}
\usepackage{siunitx}
\hypersetup{colorlinks=true,linkcolor=black,citecolor=black,filecolor=black,urlcolor=black}
\newcommand{\SHK}{\ensuremath{S _{\rm HK}}}
\newcommand{\lrhk}{\ensuremath{\log R'_{\rm HK}}}
\newcommand{\bis}{\ensuremath{{\rm BIS}}}
\newcommand{\Mearth}{\ensuremath{\rm{M}_{\oplus}}}
\newcommand{\Rearth}{\ensuremath{\rm{R}_{\oplus}}}

\newcommand{\cmps}{\ensuremath{\rmn{cm}~\rmn{s}^{-1}}}
\newcommand{\mps}{\ensuremath{\rmn{m}~\rmn{s}^{-1}}}

\title[Pinning down the mass of \mbox{Kepler-10c}]{Pinning down the mass of \mbox{Kepler-10c}: the importance of sampling and model comparison}
\author[V.~Rajpaul et al.]{V.~Rajpaul,$^{1}$\thanks{E-mail: Vinesh.Rajpaul@physics.ox.ac.uk} L.~A.~Buchhave,$^{2}$ and Suzanne Aigrain$^{1}$
\\
	$^{1}$Sub-department of Astrophysics, Department of Physics, University of Oxford, Oxford OX1 3RH, UK\\
 	$^{2}$ Centre for Star and Planet Formation, Natural History Museum of Denmark, University of Copenhagen, DK-1350 Copenhagen, Denmark
 	}
\begin{document}
\date{Accepted 2017 July 19. Received 2017 June 26; in original form 2017 May 2}
\pagerange{\pageref{firstpage}--\pageref{lastpage}} \pubyear{2017}
\maketitle
\label{firstpage}
%------------------------------------------------------------------------------------------------------------------------------
\begin{abstract}
%------------------------------------------------------------------------------------------------------------------------------
Initial RV characterisation of the enigmatic planet Kepler-10c suggested a mass of \mbox{$\sim17$~\Mearth}, which was remarkably high for a planet with radius $2.32$~\Rearth; further observations and subsequent analysis hinted at a (possibly much) lower mass, but masses derived using RVs from two different spectrographs (HARPS-N and HIRES) were incompatible at a $3\sigma$~level. We demonstrate here how such mass discrepancies may readily arise from sub-optimal sampling and/or neglecting to model even a single coherent signal (stellar, planetary, or otherwise) that may be present in RVs. We then present a plausible resolution of the mass discrepancy, and ultimately characterise Kepler-10c as having mass $7.37_{-1.19}^{+1.32}$~\Mearth, and mean density $3.14^{+0.63}_{-0.55}$~g~cm$^{-3}$.
%------------------------------------------------------------------------------------------------------------------------------
\end{abstract}
%------------------------------------------------------------------------------------------------------------------------------
%------------------------------------------------------------------------------------------------------------------------------
\section{Introduction}\label{sec:intro}
%------------------------------------------------------------------------------------------------------------------------------
%------------------------------------------
% Figure: 3 planet synthetic data fits
%------------------------------------------
\begin{figure*}
\begin{center}
\includegraphics[width=\textwidth]{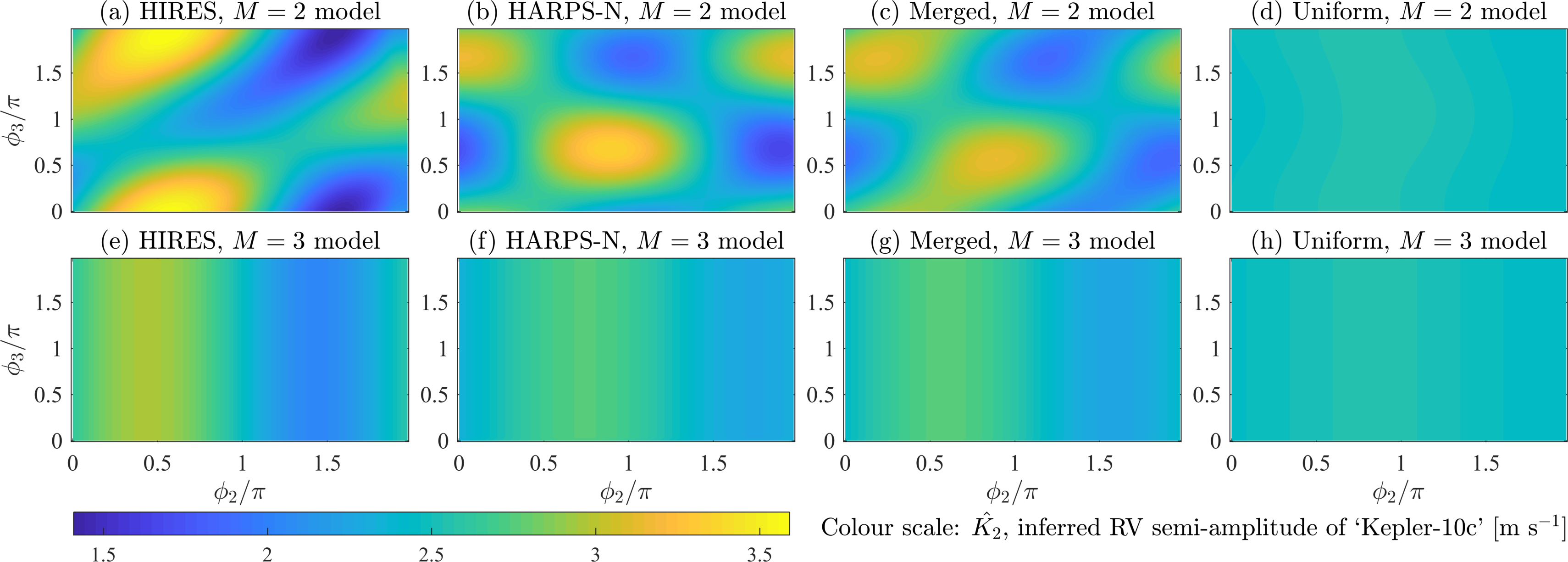}
\caption[]{ML estimates for $K_2$ based on synthetic data comprising three sine waves, with 4 different sampling patterns (left to right, corresponding to sampling patterns \ref{samplingA}--\ref{samplingD} listed on pg.\ \pageref{samplingA}) and fitted with a 2-sine model (upper panels) and a 3-sine model (lower panels). $\phi_3$ axis compressed to save space.}
\label{fig:3pfit}
\end{center}
\end{figure*}
%------------------------------------------
Kepler-10 (KOI-72; hereafter K-10 for short) is a slowly-rotating, Sun-like star that exhibits little stellar activity \citep[][hereafter D14]{D14}. It is known to host at least two planets, \emph{viz}.\ Kepler-10b, and Kepler-10c.

Stony-iron world Kepler-10b (hereafter \mbox{K-10b}) -- with orbital period $0.84$~d, radius $1.48$~$\rmn{R}_\oplus$, and mass $\sim4$~$\rmn{M}_\oplus$ -- was the first unambiguously rocky exoplanet to be discovered, and also the first super-Earth discovered around a Sun-like star \citep{batalha11}.

Kepler-10c (hereafter \mbox{K-10c}) -- with orbital period $45.29$~d, and radius $2.32$~\Rearth\ -- has proven more enigmatic. Following its discovery and statistical validation as a planet \citep{batalha11,fressin11}, D14 used 148 \mbox{HARPS-N} radial velocities (RVs) spanning two observing seasons to infer a mass of $17.2\pm1.9$~\Mearth. Given K-10c's radius, this was a striking result. Most planets with radii $2.0$--$2.5$~\Rearth\ have masses significantly lower than $17$~\Mearth, with a weighted mean mass of $5.4$~\Mearth\ \citep{weiss+marcy14}; \citeauthor{weiss+marcy14}'s empirical mass-radius relation for planets between $1.5$ and $4$~\Rearth, \emph{viz.} $M_p/\Mearth=2.69(R_p/\Rearth)^{0.93}$, predicts a mass of $5.8$~\Mearth\ for K-10c. D14 interpreted the composition of K-10c as being mostly rock by mass, and regarded the planet as being the first evidence of a class of more massive solid planets with longer orbital periods.

\citet[][hereafter W16]{weiss16} built on the work of D14, adhering closely to the techniques employed by the latter authors, but adding 72 RVs from Keck-HIRES to the analysis, resulting in a combined RV baseline of 6 years. Since it has been well established that both the HIRES and \mbox{HARPS-N} spectrometers are independently capable of accurate and precise measurement of low-amplitude planetary signals, it was a great surprise when \textcolor{black}{W16 inferred a mass for K-10c of $5.69_{-2.90}^{+3.19}$~\Mearth\ (fitted RV semi-amplitude $K_c=1.09\pm0.58$~\mps) using the HIRES RVs alone, which was incompatible with D14's estimate of $17.2\pm1.9$~\Mearth\ ($K_c=3.26\pm0.36$~\mps) using the HARPS-N RVs alone.}

W16 concluded that some additional, time-correlated signal (possibly from stellar activity or additional planets) was present and led to the discrepant mass estimates for K-10c. This claim was supported by (i) the fact that masses inferred using RVs from \emph{either} instrument were found to be time-dependent, and (ii)  $>5\sigma$ evidence for transit timing variations (TTVs) of K-10c \citep{kipping15}. W16 found that dynamical solutions including a third planet candidate, KOI-72.X, were very strongly favoured over a two-planet solution (based on Bayesian Information Criterion differentials); the TTVs and RVs were consistent with KOI-72.X having an orbital period of $24$, $71$, or $101$~d, with $101$~d being strongly favoured over the other periods. W16 inferred a likely mass of $\lesssim7$~\Mearth\ for KOI-72.X, based on the best solutions from a partial exploration of the dynamical parameter space, with the parameters of K-10b fixed. 

Even when including a third planet in their models, however, W16 were not able to reconcile the HIRES and HARPS-N masses for K-10c, so settled on a `compromise' mass for K-10c of $13.98\pm1.79$~\Mearth. We suggest the observed $3\sigma$-incompatibility between the HARPS-N and HIRES estimates for K-10c's mass points to an inadequate model under which \textcolor{black}{at} least one (if not both) of the inferred masses is incorrect, and that the true mass need not lie in the middle of the two incompatible mass posteriors.
%------------------------------------------------------------------------------------------------------------------------------
\section{Double trouble: imperfect model meets inadequately-sampled signal}\label{sec:window}
%------------------------------------------------------------------------------------------------------------------------------
To shed light directly on the effects of (i) sub-optimal sampling and (ii) inference based on an imperfect physical model, consider synthetic RV data sets $\{ (t_i,y_i)|i=1,2,\ldots,N\}$ generated as follows:
\begin{equation}
\label{eq:synthdata}
{y_i} = \sum\limits_{j = 1}^M {{K_j}\sin (\tfrac{{2\pi {t_i}}}{{{P_j}}} + {\phi _j})};
\end{equation}
$y_i$ may be interpreted as the combined RV signal at time $t_i$ due to $M$ planets on zero-eccentricity orbits around a star. For planet $j$, the associated RV amplitude $K_j$ would be determined by the planet's mass and inclination (assuming known stellar mass); $P_j$ would correspond to the planet's orbital period; and $\phi_j$ would be determined by the planet's orbital phase in some coordinate system.

Suppose we set $M=3$, and let the periods of the three mock planets be $P_1=0.84$~d, $P_2=45.29$~d, and $P_3=101.36$~d; $P_1$ and $P_2$ correspond to the known orbital periods of \mbox{K-10b} and \mbox{K-10c}, respectively, while $P_3$ corresponds to the most likely orbital period (per W16) for planet candidate KOI-72.X. We further set $K_1=K_2=2.5$~\mps, and $K_3=1.0$~\mps; the values for $K_1$ and $K_2$ are roughly the average of the RV semi-amplitudes of both \mbox{K-10b} and \mbox{K-10c} variously reported in the literature, while the value for $K_3$ is based on the most likely mass for KOI-72.X reported by W16. Lastly, instead of fixing the mutual phases of the planets, let us generate synthetic data by evaluating equation \eqref{eq:synthdata} over a grid of phases $({\phi _1},{\phi _2},{\phi _3}) \in {\Phi ^3}$ where $\Phi  = [0,0.1,0.2, \ldots ,2\pi ]$.

Now consider the problem of using such synthetic data to infer the value of $K_2$, i.e.\ the RV semi-amplitude of the second mock planet, using both a 2-planet and a 3-planet model, and with the synthetic signals sampled discretely using the following calendars:
\begin{enumerate}
\item the real HIRES observing calendar ($N=72$) for K-10; \label{samplingA}
\item the real \mbox{HARPS-N} observing calendar ($N=148$) for K-10;
\item the combined \mbox{HARPS-N}/HIRES observing calendars; and
\item 220~uniformly-spaced observations with a $\sim6$~yr baseline \textcolor{black}{(see Appendix A for more details)}. \label{samplingD}
\end{enumerate}
Thus we will produce an estimate for $K_2$, $\hat{K_2}$, using $2$ different models (of orders $M=2$ and $M=3$), $4$ different sampling patterns, and many different input values for $\phi_j$. In each case we assume that the periods and orbital phases of the planets are tightly constrained, as if performing RV follow-up of transiting planets, but that the planets' RV semi-amplitudes are not known \emph{a priori}. For the $2$-planet model we assume only the presence of the $P_1=0.84$~d and $P_2=45.29$~d planets, while for the $3$ planet model we also assume the presence of a $P_3=101.36$~d planet.

%------------------------------------------
% Figure: 3+1 planet synthetic data fits
%------------------------------------------
\begin{figure*}
\begin{center}
\includegraphics[width=\textwidth]{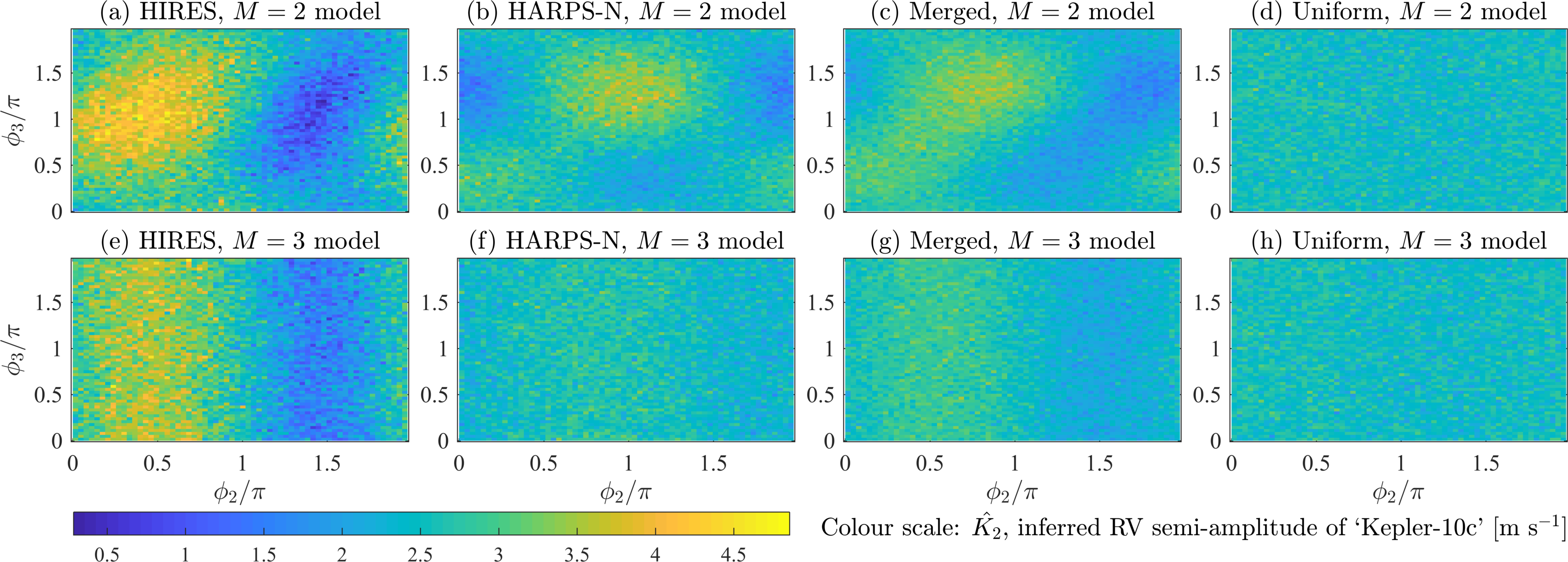}
\caption[]{As for Fig.\ \ref{fig:3pfit}, but now with \emph{noisy} synthetic data comprising four rather than three sine waves (see text on pg.\ \pageref{4sine} for more details).}
\label{fig:4pfit}
\end{center}
\end{figure*}
%------------------------------------------

We place flat priors on all free parameters, and to simplify computation, reduce the more general problem of finding posterior distributions to one of maximum likelihood (ML) estimation; our priors render ML estimates equivalent to maximum \emph{a posteriori} (MAP) estimates. As a further convenience, we assume i.i.d.\ Gaussian measurement errors on our synthetic data (with arbitrarily small, constant standard deviation), then use a standard downhill simplex algorithm (with multiple starting points) to locate ML parameters through least-squares fitting.

In Fig.\ \ref{fig:3pfit} we present the results of this fitting exercise, showing ML estimators of $K_2$, $\hat{K_2}$, for a range of values of $\phi_2$ and $\phi_3$; we fix $\phi_1$ to permit visualisation in two dimensions.\footnote{Given its short period $P_1\ll P_2<P_3$, the phase of the innermost mock planet does not have a significant effect on our inference about the properties of the other signals in our synthetic data. $K_1$ can be estimated more accurately and precisely than $K_2$, regardless of the phases of the other signals, and of the choice of observing calendar.} The following striking conclusions emerge.
\begin{enumerate}
\item When using an inadequate model (i.e.\ $M=2$) to fit the data, the inferred mass for the second mock planet is very sensitive to the mutual phases of that planet and the unobserved third planet, with $\hat{K_2}$ varying between $\sim1.5$ and $3.5$~\mps.
\item For almost all possible mutual phases of the second and the unobserved third planet, $\hat{K}_{2,\rmn{HIRES}}$ and $\hat{K}_{2,\rmn{HARPS}}$ differ\footnote{In a few cases where we chose to explore full posterior distributions for $K_2$ (rather than just find ML estimators), estimates of $K_2$ obtained with the \mbox{HARPS-N} vs.\ HIRES sampling often disagreed at a $2\sigma$ level.} by up to $\sim\pm1$~\mps\ (in the worst cases) and typically by $\sim\pm60$~\cmps. For about $60\%$ of possible phase configurations, the \mbox{HARPS-N} and HIRES observing calendars either both result in overestimation or in underestimation of $K_2$ (i.e., the true value is \emph{not} bracketed); for about $40\%$ of configurations, one calendar will lead to $\hat{K_2}>K_2$ while the other leads to $\hat{K_2}<K_2$.
\item When combining the HIRES and \mbox{HARPS-N} observations, $\hat{K_2}$ interpolates the values predicted by the separate data sets, yet may still differ from $K_2=2.5$~\mps\ by up to $\sim\pm65$~\cmps.
\item Notably, even when using the inadequate $M=2$ model but the uniform observing cadence, $|\hat{K_2}-K_2|<5$~\cmps,\;$\forall\;\phi_j$.
\item Equally notably, when using the correct $M=3$ model, the inferred mass for the second planet is relatively \emph{insensitive} to the observing calendar. {Now we find \textcolor{black}{$|\hat{K}_{2,\rmn{HIRES}}-{K}_{2}|<37$}~\cmps\ and \textcolor{black}{$|\hat{K}_{2,\rmn{HARPS}}-{K}_{2}|<22$}~\cmps}; with the combined \mbox{HARPS-N} and HIRES observations, $|\hat{K_2}-K_2|< 25$~\cmps; and with the calendar where $t_{i+1}-t_i=$ constant, $|\hat{K_2}-K_2|< 4$~\cmps,\;$\forall\;\phi_j$.
\end{enumerate}

%\phantomsection
Suppose\label{4sine} we add another signal into the synthetic RV data, with amplitude $K_4=1.0$~\mps\ and period $P_4=55$~d, as a simplistic representation of a non-evolving stellar activity signal (we could use a Gaussian process to synthesise a quasi-periodic signal instead, but such sophistication is not required for the present illustration), then repeat the exercise of trying to estimate $K_2$.%; our results are shown in Fig.\ \ref{fig:4pfit}. 

Now, neither the $M=2$ nor the $M=3$ model is adequate in that neither accounts for a fourth periodic signal present in the data. Accordingly, we find that inference about $K_2$ becomes even more sensitive to sampling. $\hat{K}_{2,\rmn{HIRES}}$ and $\hat{K}_{2,\rmn{HARPS}}$ now differ by up to $2$~\mps\ under the $M=2$ model, and by up to $1$~\mps\ under the $M=3$ model. As before, however, when using the uniform observing cadence, the resultant uniform phase coverage allows remarkably robust inference about $K_2$ to be made: e.g.,\ $|\hat{K_2}-K_2|< 7$~\cmps\;$\forall\;\phi_j$, even under the $M=2$ model.

Thus far we have established the prevalence of sizeable differences in $\hat{K_2}$ when fitting \emph{simplistic} synthetic signals with \mbox{HARPS-N} vs.\ HIRES sampling, yet  even larger differences will result when including in our synthetic data such details as photon noise, quasi-periodic stellar activity signals, instrumental noise, multiple undetected planets, planets with non-circular orbits, possible dynamical interactions between planets, and more. For example, adding to our synthetic data white Gaussian noise at a level consistent with that estimated for the \mbox{HARPS-N} dataset ($\sigma\sim2$~\mps), then repeating the previous test, results in \mbox{HARPS-N}/HIRES discrepancies for $\hat{K_2}$ of up to $3$~\mps\ under the $M=2$ model, and up to $2.3$~\mps\ under the $M=3$ model; see Fig.~\ref{fig:4pfit}.

\begin{figure}
\begin{center}
\includegraphics[width=\columnwidth]{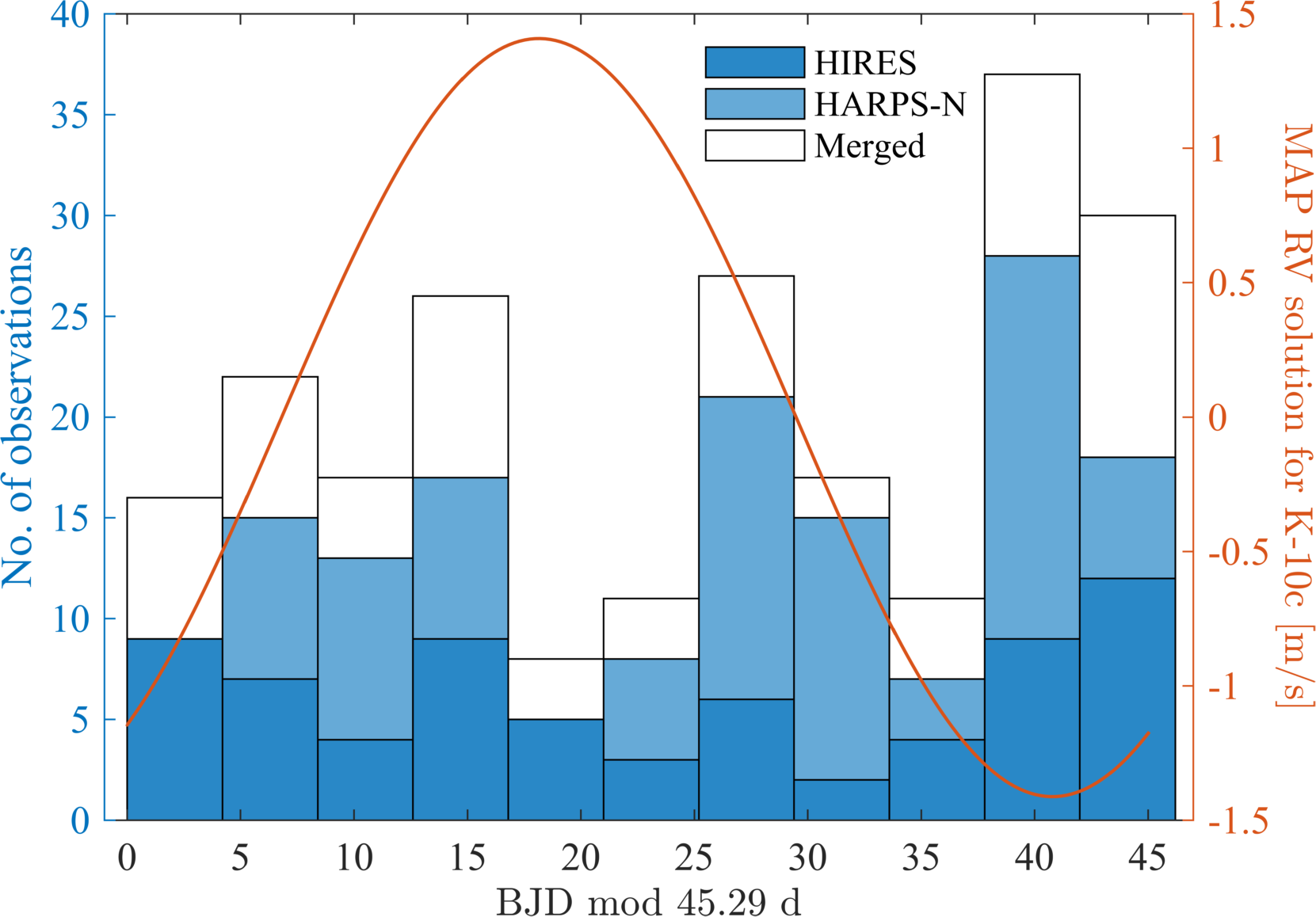}
\caption[]{\textcolor{black}{The uneven coverage of K-10c's orbital phase provided by existing HARPS-N and HIRES observations. \textcolor{black}{(This representation does not by itself indicate whether the sampling would lead to mass under- or over-estimation; this would require accounting for constructive or destructive interference between K-10c's signal and all other signals in the RVs.)}}}
\label{fig:phasebins}
\end{center}
\end{figure}

The upshot is that using an inadequate physical model, and/or suboptimal sampling, can lead to incorrect conclusions about the masses even of planets whose other properties are well constrained -- and even when we have hundreds of RVs at our disposal. Moreover, through our choice of real sampling patterns, and realistic values for $K_j$ and $P_j$, we have provided a plausible explanation for why W16 obtained discrepant masses for \mbox{K-10c} using HIRES vs.\ \mbox{HARPS-N} RVs. Specifically, the real K-10 RVs likely contained not only \mbox{K-10b} and \mbox{K-10c}'s signals, but one or more other coherent signals (KOI-72.X, a stellar signal, etc., as indeed adduced by W16) which interfered constructively or destructively with the signals of the known planets. \textcolor{black}{The sub-optimality of the phase coverage is easily checked by phasing the HIRES or HARPS-N observation times to the orbital period of \mbox{K-10c}; see Fig.\ \ref{fig:phasebins}}. In principle, accounting for the other signals jointly with those of the known planets (i.e., using a more appropriate physical model), and/or obtaining more observations to provide more complete phase coverage of K-10c's signal, could have mitigated the discrepancy.

%-------------------------------------------------------------------------------------------------------------------------------
\section{Reconciling the mass estimates}\label{sec:origin}
%-------------------------------------------------------------------------------------------------------------------------------

%------------------------------------------
% Figure: phased observations
%------------------------------------------
%------------------------------------------

%------------------------------------------
\begin{table*}
\centering
\caption{Summaries of marginal posteriors for selected planet \textcolor{black}{and GP} parameters from our favoured model (three planets plus correlated noise). The planet parameters are the same as those from W16\textcolor{black}{, while the GP parameters are as defined in R15; $K_\rmn{GP}=\sqrt{V_\rmn{r}^2+V_\rmn{c}^2}$} \textcolor{black}{may be interpreted as the GP RV semi-amplitude; $P$ is an overall period; $1/\lambda_\rmn{p}$ defines the harmonic complexity of the GP (behaviour is sinusoidal for $\lambda_\rmn{p}\gg1$); and $\lambda_\rmn{e}$ is the time scale over which the GP signal evolves.} Periapsis passage times are omitted: for K-10b and K-10c, these were effectively fixed in our models \emph{a priori} via known transit times, while a periapsis passage time for non-transiting planet candidate KOI-72.X could not be well constrained, given its apparently-circular orbit.}
\begin{tabular}{llrlrlrl}
\hline
&  & \multicolumn{2}{c}{HIRES} & \multicolumn{2}{c}{HARPS-N} & \multicolumn{2}{c}{Merged} \\ %\hline

Parameter & Units & \multicolumn{1}{r}{Median} & \multicolumn{1}{l}{$\pm\sigma$}       & \multicolumn{1}{r}{Median} & \multicolumn{1}{l}{$\pm\sigma$}        & \multicolumn{1}{r}{Median} & \multicolumn{1}{l}{$\pm\sigma$}  \\ \hline

\rule{0pt}{2.5ex} $K_\rmn{b}$   & \mps  & 2.39  		& $_{-0.28}^{+0.30}$	& 2.33  & $\pm0.16$  & 2.32  & $_{-0.18}^{+0.21}$ \\
$P_\rmn{b}$   				& days  & $0.83748$  	& $\pm0.00003$  		& 0.837501  & $\pm0.000005$  & 0.837501  & $_{-0.000004}^{+0.000005}$ \\
$\sqrt{e_\rmn{b}} \cos\omega_\rmn{b}$ 		& -     & $0.000$ & $\pm0.003$ 	& $0.000$  & $\pm0.003$  & $0.000$  & $\pm0.004$ \\
$\sqrt{e_\rmn{b}} \sin\omega_\rmn{b}$ 		& -     & $0.000$ & $\pm0.003$ 	& $0.000$  & $\pm0.003$  & $0.000$  & $\pm0.004$ \\
 $m_\rmn{b}$   & \Mearth  & 3.33  		& $_{-0.42}^{+0.40}$	& 3.25  & $_{-0.23}^{+0.22}$  & $3.24$  & $\pm0.28$ \\
 $\rho_\rmn{b}$   & g~cm$^{-3}$  & 5.65  		& $_{-0.85}^{+0.94}$	& $5.51$  & $_{-0.64}^{+0.73}$  & $5.48$  & $_{-0.68}^{+0.78}$ \\
\hline

\rule{0pt}{2.5ex}  $K_\rmn{c}$  & \mps  & $1.27$  		& $_{-0.35}^{+0.42}$  & $1.64$  & $_{-0.34}^{+0.42}$  & $1.41$  & $_{-0.23}^{+0.25}$ \\
$P_\rmn{c}$   				& days  & 45.2948 		& $\pm0.0008$  	& $45.2940$ & $_{-0.0007}^{+0.0008}$  & $45.2946$ & $\pm0.0008$ \\
$\sqrt{e_\rmn{c}} \cos\omega_\rmn{c}$ 		& -     & $0.1$  & $\pm0.2$  & $0.0$  & $\pm0.1$  & $0.0$  & $\pm0.1$ \\
$\sqrt{e_\rmn{c}} \sin\omega_\rmn{c}$ 		& -     & $0.0$  & $\pm0.2$  & $0.1$  & $\pm0.1$  & $0.0$  & $\pm0.1$ \\
 $m_\rmn{c}$   & \Mearth  & $5.87$  		& $_{-1.82}^{+2.20}$	& $8.59$ & $_{-1.79}^{+2.19}$  & $7.37$  & $_{-1.19}^{+1.32}$ \\
 $\rho_\rmn{c}$   & g~cm$^{-3}$  & 2.50  		& $_{-0.78}^{+0.98}$	& $3.66$  & $_{-0.80}^{+0.98}$  & $3.14$  & $_{-0.55}^{+0.63}$ \\
\hline

\rule{0pt}{2.5ex} $K_\rmn{X}$   & \mps  & $1.30$  & $_{-0.45}^{+0.51}$  		& $0.84$  & $_{-0.14}^{+0.16}$  & $0.85$  & $_{-0.14}^{+0.24}$ \\
$P_\rmn{X}$   				& days  & $102$ & $_{-7}^{+8}$  		& 101 & $_{-5}^{+6}$  & $102$ & $\pm1$ \\
$\sqrt{e_\rmn{X}} \cos\omega_\rmn{X}$ 		& -     & $-0.1$  & $\pm0.2$  		& $-0.1$  & $\pm0.1$  & $-0.1$  & $\pm0.1$ \\
$\sqrt{e_\rmn{X}} \sin\omega_\rmn{X}$ 		& -     & $-0.1$  & $\pm0.2$  		& $0.0$  & $\pm0.1$  & $0.0$  & $\pm0.1$ \\
$m_\rmn{X}$   & \Mearth  & 8.93  		& $_{-3.15}^{+3.50}$	& 5.80 & $_{-1.03}^{+1.20}$  & $5.90$  & $_{-1.01}^{+1.70}$ \\
\hline
\rule{0pt}{2.5ex}
\textcolor{black}{$K_\rmn{GP}$}  	& \mps  & $0.09$ & $_{-0.06}^{+0.22}$  		& $1.46$ & $\pm0.17$  & $1.68$ & $\pm0.25$ \\
\textcolor{black}{$P$}    				& days  & 63 & $\pm10$  		& 55 & $\pm1$  & $55.5$ & $\pm0.8$ \\
\textcolor{black}{$\lambda_\rmn{p}$} 	& -     & $1.3$  & $_{-0.3}^{+0.6}$  		& $0.33$  & $_{-0.02}^{+0.04}$  & $0.32$  & $_{-0.01}^{+0.02}$ \\
\textcolor{black}{$\lambda_\rmn{e}$ }		& days     & $330$  & $\pm100$  		& $86$  & $\pm4$  & $90$  & $\pm6$ \\

\hline

\end{tabular}%
\label{tab:bestmodel}%
\end{table*}%
%------------------------------------------
We noted $>3\sigma$ evidence for linear correlations ($\rho\sim30\%$) between the published HARPS-N RVs and (i) \lrhk\ index and (ii) \bis\ (bisector inverse slope) measurements; we did not find any similarly-significant correlations in the HIRES RVs.

Whereas the models of D14 and W16 did not accommodate possible stellar activity signals in the RVs, we used the Gaussian process (GP) framework of \citet[][hereafter R15]{Rajpaul2015} to model jointly all available RV, \lrhk\ (in the case of HARPS-N) or \SHK\ (in the case of HIRES), and \bis\ time series, for a total of $660$~datapoints. As in R15, we adopted a quasi-periodic covariance kernel, and non-informative priors were placed on all GP hyper-parameters. GP amplitude parameters were also constrained to be smaller than the total variation seen in a given time series, and of the overall GP period we required $P>20$~d (based on D14's lower limits on K-10's stellar rotation period). We additionally allowed at least $2$ but up to $5$ possible planetary signals in the RVs, modelled with Keplerian functions. We constrained the periods and periapsis passage times of two of the Keplerians to be consistent with the most precise values inferred from K-10b and K-10c's transits \citep{morton16,holczer16}, but left the other parameters free, with priors identical to those in W16's eccentric $2$-planet model. We adopted analogous uninformative priors for all parameters of the additional possible planets, ensuring only that planet periods did not overlap. Finally, we used the \textsc{MultiNest} nested-sampling algorithm \citep{multinest2008,multinest2009,multinest2013} to obtain a full joint posterior distribution for each model's parameters (and marginal posteriors for parameters of interest), and to compute a Bayesian evidence ($\mathcal{Z}$) for each model.

Of the numerous models we considered, we found only one in which estimates for \emph{all} planet parameters were consistent within $1\sigma$ between the HARPS-N, HIRES, and merged data sets: \emph{viz.} a model including three planets, all with orbits consistent with circular, plus correlated noise. Significantly, this model was also favoured over others by Bayesian model comparison tests, and the period for the third planet in our model was $102\pm1$~d: in accord with the W16's favoured period for KOI-72.X (based on both analytical considerations and dynamical modelling), despite us not including this as prior information in our model. We summarise the marginal posteriors for this favoured model's planet parameters in Table \ref{tab:bestmodel}; masses (for all three planets) and mean densities (for the transiting planets) were derived using the same stellar mass and planet radii as in W16. 

Additionally, we note the following. First, for the HIRES, HARPS-N and merged data sets, 3-planet models were strongly favoured over 2-planet models ($\Delta\ln\mathcal{Z}>10$), which were in turn favoured over 4- and 5-planet models. Secondly, we obtained consistent parameters for all planets when splitting either the HIRES or HARPS datasets in two; presumably W16 found discrepant results because neither a third planet nor a nuisance signal model was included when performing the same test. Thirdly, a zero-amplitude GP component was favoured for the HIRES RVs ($\Delta\ln\mathcal{Z}\sim3$), whereas a non-zero GP amplitude was favoured ($\Delta\ln\mathcal{Z}\gtrsim 10$) for the HARPS-N and merged RVs; the latter two cases suggested a GP period of $P=55\pm1$~d.

We interpreted the third finding as evidence of the HARPS-N RVs being confounded by at least one semi-coherent though not strictly periodic, $\gtrsim1$~\mps\ nuisance signal that cannot not be ascribed to a planet (for which a simpler Keplerian model would have sufficed, rather than a GP; a planetary signal should also have been simultaneously present in the HIRES RVs). Given the correlation observed between HARPS-N RVs and activity indicators, at least part of this signal could be due to stellar activity; it is unclear, though, whether $P=55\pm1$~d corresponds to a stellar rotation period. An instrumental component to the signal also cannot be ruled out.\footnote{\textcolor{black}{The small posterior uncertainty of $\pm1$~d may simply indicate that $55$~d is the only GP period that does a reasonable job of modelling some (possibly complex) combination of nuisance signals. Regardless, given the similarity of the $55$~d period to K-10c's orbital period, the nuisance and planet signals interfere strongly over time scales of several months (an envelope with period $248$~d would be predicted if the nuisance signal were sinusoidal).} } Either way, it appears that when \emph{not} accounting for a correlated nuisance signal, and an apparent third planet (KOI-72.X), the amplitude of the signal ascribed to K-10c is forced to inflate artificially to absorb some of this appreciable variability in the discretely-sampled RV signal.

%------------------------------------------
% Figure: mass-radius diagram
%------------------------------------------
\begin{figure*}
\begin{center}
\includegraphics[width=\textwidth]{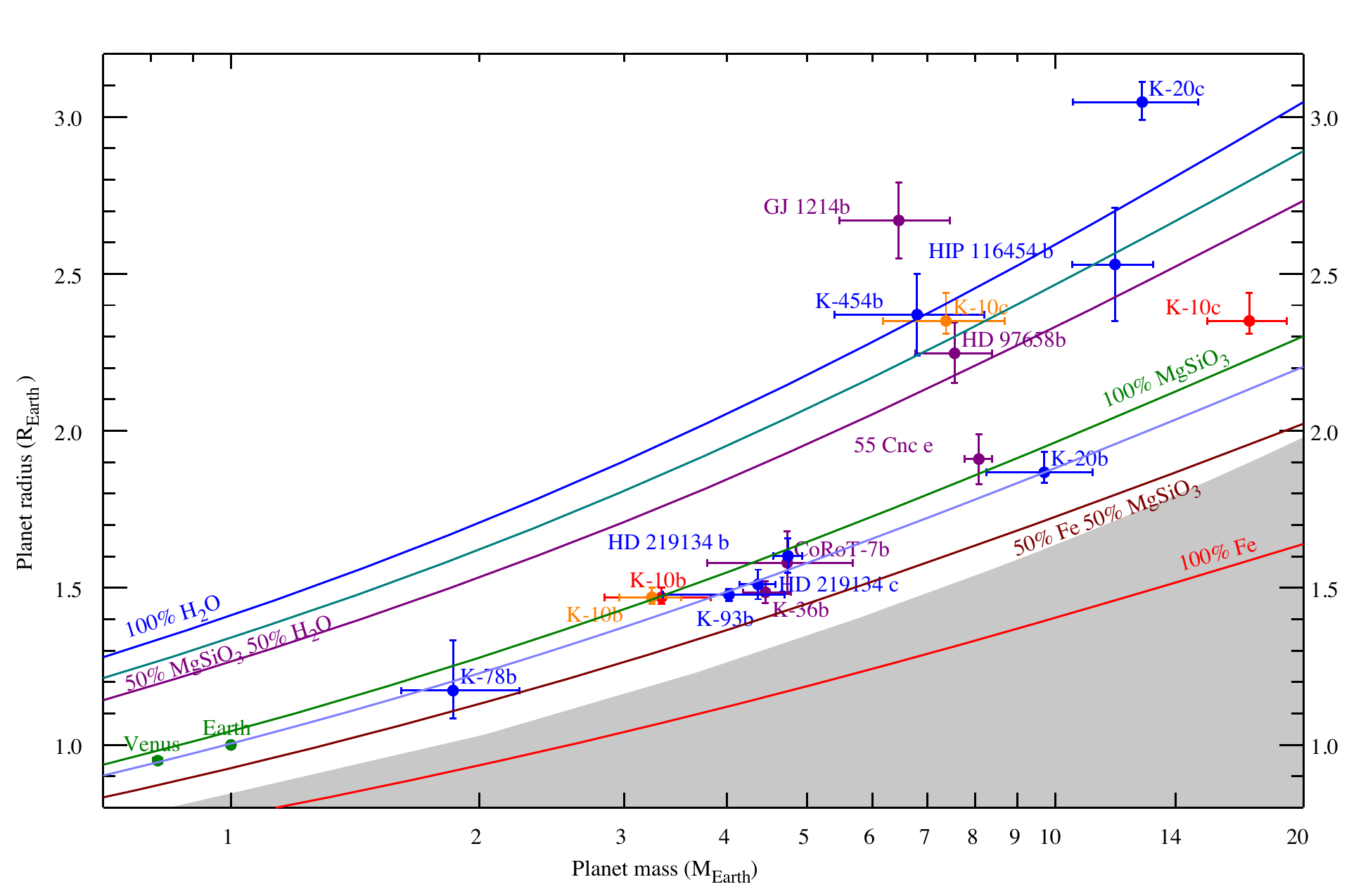}
\caption[]{\textcolor{black}{Mass-radius relation for planets smaller than $3.2$~\Rearth, and mass determinations better than $20\%$~precision. The shaded region denotes where planets would have an iron content exceeding the maximum value predicted from collisional stripping models. Solid curves are theoretical models for planet with a composition consisting of 100\% H$_2$O (blue), 25\% silicate and 75\% H$_2$O (teal), 50\% silicate and 50\% H$_2$O (magenta), 100\% silicate (green), 70\% silicate and 30\% iron, consistent with an Earth-like composition (light blue), 50\% silicate and 50\% iron (brown), and 100\% iron (red)  \citep{zeng2013}. Blue points indicate planets with masses measured using the HARPS-N spectrograph, while purple points are from other sources. The red and orange points correspond to the confirmed K-10 planets, as characterised by D14 and this paper, respectively.}}
\label{fig:mass-radius}
\end{center}
\end{figure*}
%------------------------------------------

Our mass and mean density estimates for K-10b are consistent with those of D14 and W16. Our mass estimate for K-10c ($7.37_{-1.19}^{+1.32}$~\Mearth), however, is significantly lower than those from D14 and W16 ($17.2\pm1.9$~\Mearth\ and $13.98\pm1.79$~\Mearth, respectively); accordingly, we also infer a significantly lower mean density of $\rho_\rmn{c} = 3.14^{+0.63}_{-0.55}$~g~cm$^{-3}$. This implies a composition which is either consistent with a low-density solid planet with a significant fraction of volatiles in the form of e.g.\ water or methane, or a planet with a dense core and an extended gaseous envelope. K-10c would thus join a region of parameter space in the mass-radius diagram occupied by a number of other exoplanets with radii between 2.0 and $2.5~\Rearth$ that have similar mean densities to K-10c; \textcolor{black}{see Fig.\ \ref{fig:mass-radius}}. 

Finally, our inferred mass of $5.90_{-1.01}^{+1.70}$~\Mearth\ for KOI-72.X is compatible with W16's point estimate of $\sim7$~\Mearth, though it remains to be established whether this is a genuine planet. We used the parameters from our Keplerian solutions as inputs to numerical N-body integrations using \textsc{TTVFast} \citep{deck14}; we noted that the maximum difference between the RVs from a full dynamical simulation and from our Keplerian solution was of the order $~1$~\cmps\ over $101$~d. As this is two orders of magnitude below the RV noise floor, we concluded that full dynamical modelling would have yielded no constraints beyond those already derived by W16.

%Finally, our inferred mass of $5.90_{-1.01}^{+1.70}$~\Mearth\ for planet candidate KOI-72.X is compatible with W16's point estimate of $\sim7$~\Mearth, though it remains to be established whether this is a genuine planet. We used the parameters from our Keplerian solutions as inputs to a series of numerical N-body integrations using \textsc{TTVFast} \citep{deck14}; we noted that the maximum difference between the RVs from a full dynamical simulation and from our Keplerian solution (which neglected dynamical interactions between planets) was of the order $~1$~\cmps\ over $101$~d. As this difference is two orders of magnitude below the RV noise floor, we concluded that full dynamical modelling would have yielded no constraints beyond those already derived by W16.

%---------------------------------------------- --------------------------------------------------------------------------------
\section{Instrumental considerations}\label{sec:other}
%-------------------------------------------------------------------------------------------------------------------------------

W16 detrended the HIRES RVs by removing correlations between RVs and instrumental parameters, RV uncertainties, and spectrum signal-to-noise ratio. The RVs published in W16 are these \emph{detrended} RVs; the published RV uncertainties also already have jitter applied. We obtained both the pre-detrending RVs and the uncertainties without jitter from Weiss (\emph{pers.\ comm.}), and re-ran the analyses described in Section \ref{sec:origin}. As before, we ended up favouring a $3$-planet plus correlated noise model strongly over all competing models, and the posterior distributions for the parameters of all three Keplerians were consistent {($<1\sigma$)} with those obtained when using the detrended HIRES RVs.

To explore the possibility of the \mbox{HARPS-N} data reduction pipeline contributing to the discrepancy, we applied a novel, template- and mask-free approach we are developing (paper \emph{in prep.}) for extracting RVs from observed spectra.  We model each observed spectrum non-parametrically, with shifts between all possible pairs of spectra included as parameters in the modelling (in addition to possible telluric, stellar activity, and instrumental effects). Interestingly, we found that when modelling HARPS-N RVs extracted with our own pipeline vs.\ the \mbox{HARPS-N} pipeline, our inferred RV semi-amplitude for K-10b was unchanged, but we reliably inferred $K_\rmn{c}<2$~\mps\ even \emph{without} a correlated noise (GP) component in our model. This suggests the possibility that at least part of the signal confounding K-10c's signal might be instrumental rather than stellar (and would explain why the same nuisance signal is absent from the HIRES RVs); given the preliminary nature of our pipeline, however, further investigation is required.

%------------------------------------------------------------------------------------------------------------------------------
\section{Discussion and conclusions}\label{sec:discuss}
%------------------------------------------------------------------------------------------------------------------------------
Previous studies \citep[e.g.][]{dawsonfab2010} have explored the impact of irregular time sampling on planet \emph{period} estimation; here we have demonstrated that a failure to account for one or more coherent signals (whether of stellar, planetary, or instrumental origin) in RV data, and/or uneven phase coverage, can confound attempts to infer the \emph{masses} of planets with known periods. We used synthetic data with sampling based on real observations to demonstrate how such difficulties could arise when characterising planets in a system analogous to K-10; tests such as the ones we presented may readily be applied to other systems, to test the sensitivity of planet characterisation to sampling and model selection.

By accounting for a time-correlated (stellar or instrumental) signal present in the HARPS-N K-10 RVs, as well as a likely third planet in the system, we were able to achieve full consistency between the Keplerian solutions for the HIRES, HARPS-N, and combined RVs. The third planet included in our model has properties consistent with K-10c's TTVs; and although our model is more complex than the one used by W16 to model the RVs, it was nevertheless favoured over simpler models in Bayesian model comparison testing. While our proposed resolution of the K-10c mass discrepancy is a plausible one, it appears that (many) more RVs will be required for a definitive characterisation of the K-10 system.

Whereas W16 suggested a strategy of employing a long observing baseline compared to time-correlated noise influences, we suggest it's also important to focus on obtaining more complete \emph{phase coverage} of the relevant signals. As we demonstrated in Section \ref{sec:window}, good phase coverage can permit robust inference about known planets, even when using a demonstrably-inadequate physical model. While uniform cadence might not be feasible or desirable, e.g.\ to avoid aliasing, a long observing baseline and approximately-uniform cadence would lead to good phase coverage even of planets with unknown orbital periods \textcolor{black}{(see Appendix A for more details)}. And while W16 suggested that a long baseline would help to average out spurious signals that may arise from stellar activity, we suggest it is strongly preferable to \emph{model} these nuisance signals, as it is difficult to know \emph{a priori} how these nuisance signals might interfere with signals of interest. Baselines and cadence aside, it seems all but essential to implement a variety of physical models (to account for varying numbers of possible planets, nuisance signals, etc.), and to compare systematically the evidence for the competing models. 

Our findings may also have relevance to archival RV data sets, and indeed, this is not the first example of a system where inference has turned out to be extremely sensitive to both sampling and model choice, despite the availability of a large number of RVs \citep{rajpaul16}. Then again, K-10 might have been a relatively pathological case; as W16 noted, there were various hints (TTVs, K-10c mass discrepancy, etc.) that existing characterisations of the system were inadequate. Looking to the future, with a new generation of RV spectrographs with expected precisions of $10$~\cmps\ soon to come online, optimised sampling strategies and careful model selection will clearly both be essential if these spectrographs are to be used for accurate characterisation of small planets, especially those in potentially multi-planet systems. \textcolor{black}{Moreover, it would be prudent to coordinate observations made by different teams with different telescopes, to minimise `redundant' observations that do not contribute to improved coverage of a given planet's orbital phases.}

%------------------------------------------------------------------------------------------------------------------------------
\section*{Acknowledgments}
%------------------------------------------------------------------------------------------------------------------------------

The authors thank Lauren Weiss for many useful discussions, and for sharing unpublished data (Mt.\ Wilson \SHK\ measurements and HIRES RVs before trend removal) which we used to explore alternative explanations for the \mbox{K-10c} mass discrepancy. \textcolor{black}{The authors also thank the anonymous referee for helpful feedback.} V.~R.\ is grateful to Merton College and the National Research Foundation of South Africa for providing financial support for this work.
%------------------------------------------------------------------------------------------------------------------------------
\bibliography{biblio}
\bibliographystyle{mnras}
%------------------------------------------------------------------------------------------------------------------------------
\appendix
%------------------------------------------------------------------------------------------------------------------------------
\section{On uniform cadence and phase coverage}
%------------------------------------------------------------------------------------------------------------------------------

To illuminate the connection between uniform observing cadence and uniform observational coverage of a planet's orbital phase, consider first the simple case where a planet with known orbital period $P$ is observed $N$ times over the course of one full orbit.

By the assumption of uniform cadence, and through suitable choice of origin for the time coordinate, the observation times may be written $\{ \tfrac{P}{N},\tfrac{{2P}}{N}, \ldots ,P\}$. If we now phase the observation times to the period of the planet, by computing the observation times modulo $P$, i.e.\ the remainder after division by $P$, we find
$\{ \tfrac{P}{N},\tfrac{{2P}}{N}, \ldots ,P\} \bmod P=P\left[ {\{ \tfrac{1}{N},\tfrac{2}{N}, \ldots ,1\} \bmod 1} \right] = P\left[ {\{ 0,\tfrac{1}{N}, \ldots ,1 - \tfrac{1}{N}\} } \right]$. Thus we see, trivially, that the phased observations will cover the planet's orbit uniformly; a $\sqrt{N}$-bin histogram (say) of the phased observations should contain $\sqrt{N}$ observations in each bin.\footnote{In this paper, we argued that uniform phase coverage was \emph{preferable} to very uneven phase coverage for constraining Keplerian signal amplitudes; this does not mean, however, that uniform phase coverage is necessarily optimal, as it might be possible to obtain even tighter constraints on amplitudes e.g.\ by sampling preferentially where the signals are largest. Further work is required to investigate these possibilities for Keplerian signals.}

Suppose now we extend this to the case of observations covering $M\in \mathbb{N}$ full orbits, again with uniform cadence, and $N$ observations per orbit. The $MN$ observation times are now $P\{ \tfrac{1}{N},\tfrac{2}{N}, \ldots ,1, \ldots ,M\}$, which we can write as $P\cdot\bigcup\limits_{i = 1}^M {{T_i}}$, where $\bigcup$ indicates set union, and the $T_i$ are multisets\footnote{A multiset generalises the concept of a set to allow multiple instances of the same element. For example, $\{1,1,2\}$ and $\{1,2\}$ are different multisets though they are the same set.} such that $T_1=\{ \tfrac{P}{N},\tfrac{{2P}}{N}, \ldots ,P\}$, and $T_{i+1}=T_{i}+1$. If we phase these observation times to the period $P$, we find

\begin{equation}
\left[ {P \cdot \bigcup\limits_{i = 1}^M {{T_i}} } \right]\bmod P = P\left[ {\bigcup\limits_{i = 1}^M {{T_i}} \bmod 1} \right] = P \cdot \bigcup\limits_{i = 1}^M {{T_1}},
\end{equation}
since $T_j \mod 1 = T_1$. Thus the phased observation times will simply be the uniformly-distributed observation times for the first orbit, but repeated $M$ times, so that a $\sqrt{N}$-bin histogram should contain $M\sqrt{N}$ observations in each phase bin.

Now we can consider two generalisations of the above cases. First, suppose the planet's orbital period is \emph{not} known, but that the observing cadence is uniform, and that the observing baseline is significantly longer than a single orbital period. We can then truncate the observation times to just the $M\in\mathbb{N}$ full orbits covered by the observations, and apply the above arguments to show that this large subset of observations will be evenly-distributed when phased to the orbital period of the planet. The discarded observations covering an incomplete orbit \emph{will} contribute to slight non-uniformity, but the effect should be small since $M\gg1$. Importantly, the approximately uniform phase coverage will apply to \emph{all} observed planets, provided all have periods significantly shorter than the observing baseline.

Second, suppose the observational cadence is \emph{not} strictly uniform. If the uniformity of orbital phase coverage is studied by making a histogram of the phased observation times, provided the deviations from uniform cadence are significantly smaller than the histogram bin widths, the number of items in each bin should not change compared to the case of strictly-uniform cadence.\footnote{One could derive criteria under which this approximate-uniformity would be ensured, although it is much easier to verify numerically that synthetic observation times with approximately-uniform cadence phased to an unknown period will lead to approximately uniform phase coverage, provided the observing baseline is much longer than the period.}

We have shown that (approximately) uniform observing cadence will lead to (approximately) uniform orbital phase coverage. Note that the latter, however, does not require or imply the former. As a counterexample, consider again the first case of observations made at times $\{ \tfrac{P}{N},\tfrac{{2P}}{N}, \ldots ,P\}$ for a planet with orbital period $P$. It is easy to verify that translating any number of observation times by $nP$ where $n\in\mathbb{Z}$ will lead to identical phase coverage, even though the new observation times will no longer themselves be uniformly-spaced in time.

In case (iv) considered in Section 2, i.e.\ `220 uniformly-spaced observations with a $6$~yr baseline', the observation times were for simplicity spaced $6~$yr$/220\sim10$~d apart, with no retrospective consideration given to weather, K-10's visibility, etc.\ on the chosen dates. Given the long baseline and the relatively short (known) orbital period of K-10c, i.e.\ $45.29$~d, it would be straightforward to use the above considerations to devise a more realistic observing programme that ensured equivalent phase coverage.

More generally, it should be noted that the above considerations do not necessarily apply to the case of planets with long orbital periods, say of the order several months or more. If a planet's orbital period is long but known, it should still be straightforward to plan observations, possibly spanning more than one observing season, to provide uniform phase coverage. If the orbital period of a planet is unknown \emph{and} comparable to the baseline of observations, however, it will not be possible to guarantee uniform phase coverage \emph{a priori}. Finally, uniform phase coverage is not possible with an observing baseline shorter than a planet's orbital period.

\bsp
\label{lastpage}
%------------------------------------------------------------------------------------------------------------------------------
\end{document}